\documentclass[floatfix,aps,pre,showpacs,floats,preprint]{revtex4}
\usepackage{graphicx}
\usepackage{amsfonts}
\begin{document}
\def\E{{\bf E}}

\def\P{{\bf P}}

\def\D{{\bf D}}

\def\r{{\bf r}}

\def\dV{{\rm d^3}{ r}\;}

\def\qq{{\bf q}}

\def \div{ {\nabla \cdot }\, } 

\def \grad{ {\nabla}\, } 

\def \curl{ {\rm curl}\, }

\def\p{{\bf p}}

\def\rhat{\hat {\bf r}}

\def \dv{\; d^3\rr}

\title{Fluctuation-induced interactions between dielectrics in general
  geometries} \author{S. Pasquali, A. C. Maggs} \affiliation{Laboratoire
  de Physico-Chime Th\'eorique, Gulliver CNRS-ESPCI 7083, 10 rue Vauquelin, 75231
  Paris Cedex 05, France.}
\begin{abstract}
  We study thermal Casimir and quantum non-retarded Lifshitz interactions
  between dielectrics in general geometries. We map the calculation of the
  classical partition function onto a determinant which we discretize and
  evaluate with the help of Cholesky factorization.  The quantum partition
  function is treated by path integral quantization of a set of interacting
  dipoles and reduces to a product of determinants.  We compare the
  approximations of pairwise additivity and proximity force with our numerical
  methods. We propose a ``factorization approximation'' which gives rather
  good numerical results in the geometries that we study.
\end{abstract}
\pacs{05.10.-a  % computation methods in stat physics
41.20.Cv % Electrostatics
}
\maketitle

\section{Introduction}
Much is known about fluctuation-induced interactions between bodies.  One
distinguishes two parts to the interaction \cite{Parsegian,Parsegian2,Daicic}:
The first is quantum in nature, and corresponds to a multi-center
generalization of London dispersion forces.  The second, the so called thermal
Casimir effect, has as its origin the thermal excitation of polarization modes
in dielectrics, giving rise to temperature dependent forces. Together they
form the multi-body generalization of the long-ranged part of the van der
Waals interaction.  An alternative, but equivalent vision comes from
associating these forces to the energy and entropy of fluctuating
electrodynamic fields.  A number of sophisticated theoretical techniques have
been applied to the calculation of this interaction, but only the simplest of
geometries are analytically tractable to exact solution, for instance planar
surfaces \cite{NinhamJCP}, spheres or cylinders.

More complicated physical situations are generally studied with perturbation
theory, or in exceptional cases with non-perturbative methods adapted to
individual geometries \cite{Emig}.  Such methods, that keep into account the
totality of the interaction, are usually developed on the basis of Lifshitz
theory for dielectrics.  Analytic methods are typically based on an
approximation around one of the few solvable problems. These range from the
simple proximity force approximation, to a semi-classical interactions
approximation \cite{Spruch}, to the multipole expansion \cite{Duplantier}, to
the more recent and sophisticated ray optics approximation \cite{Jaffe3}.  The
main problem with these methods is they are limited when the system under
study departs too far from the unperturbed system. Given the limitations of
all analytic methods, which are in practice only capable of studying very
regular geometries, numerical approaches have been developed. These methods
have been limited both by storage space required and by long computational
times, making it possible to study only two-dimensional systems, or
three-dimensional systems translationally invariant in one direction
\cite{Buesher,Gies}.  It is only very recently that more powerful methods have
been developed based on the numerical calculation of the Maxwell stress tensor
\cite{MIT}.  The numerical approach we present in this paper is complementary
to this last method, but it is based on somewhat different theoretical
grounds.  We will explain in the discussion how the two methods compare in
detail.

We note, as an aside, that so called ``atomistic modeling'' of materials, as
performed with most molecular dynamics codes, with the aim of understanding
micro-mechanical response, conformations of macromolecules or interfaces,
often uses assumptions such as pairwise additivity of interactions
\cite{Hofer} parameterized with phenomenological Lennard-Jones potentials.
These methods neglect retardation, screening of the classical interaction by
ions and multibody effects. It is of clear interest to develop methods which
will enable one to have a better quantitative understanding of such effects in
both soft and hard condensed matter physics.  More sophisticated (and even
more costly) quantum simulation methods based on local density functionals
exsist, but are known to miss long-ranged dispersion interactions completely
\cite{Ardhammar}.

In this paper we want to focus on important geometries which are clearly
beyond perturbative study.  One simple example is the case of a tip near a
structured surface, where the curvature of the surface makes most
approximation methods ineffective, unless the dielectric contrast is very weak
\cite{Golestanian}.  This paper has as its principle aim the generalization of
a recent paper \cite{long} which used direct diagonalization of a large matrix
in order to calculate the free energy of fluctuating dielectrics in the
classical (thermal Casimir) limit. The methods in \cite{long} were rather
limited, only very small three dimensional systems could be studied.  Quantum
effects were neglected entirely. In this paper we use more sophisticated
factorization techniques which allow us to treat finer discretizations of
physical systems.  We will, in addition, show how to introduce quantum
mechanics into our formalism and use it to study the full, non-retarded
interaction between two dielectric bodies.  {In order to compare with our
  numerical method, we analyze the performance of two simple approximation
  methods}:
%Using our numerical methods we
%analyze two simple approximate methods: 
pairwise additivity, and the  proximity
force approximation. We also introduce an approximation based on the
factorization of the geometric and material properties that works surprisingly
well for the geometries that we study.

We begin by formulating the theory of classical fluctuating dielectrics. In
this we prefer a formulation of the classical interaction in terms of
the true microscopic fluctuating field, the polarization, rather than the
electric field and electric displacement. We then show how to efficiently
factorize the resulting quadratic forms and apply our formalism to calculate
the free energy of interaction between a tip and an indentation.  We then
generalize our microscopic energy functional in order to consider the quantum
fluctuations of a dielectric and evaluate the non-retarded interaction between
a tip and a structured surface.

\section{Classical Fluctuating Dielectrics}
\label{theory} 

We begin by evaluating the classical thermal interaction, since most of the
technical difficulties of discretization and factorization are already
present, before showing how to treat the quantum case.  We start with the
energy of a heterogeneous dielectric system \cite{Marcus} written in terms of
the polarization density $\p$.  We note that formulation in terms of the
polarization field is also much more convenient for the generalization to
scale dependent dielectric effects \cite{ralf}, which are particularly
important in water based systems.

The energy of a linear dielectric has two contributions \cite{Marcus}, firstly
a long-ranged Coulomb energy for the induced charged density $\rho_i= -\div
\p$ and secondly a local contribution which depends on the local electric
susceptibility $\chi_0({\bf r})$.
\begin{equation}
  U_p = \int\;
  {\div {\bf p}({\bf r})  \div  {\bf p} ({\bf r}')  
\over 8 \pi |{\bf r}-{\bf r}'|}
 \dV \dV'
  + \int \; {{\bf p}^2({\bf r}) \over 2 \chi_0({\bf r})} \dV.
  \label{marcus}
\end{equation}
We note that the dielectric constant $\epsilon({\bf r})= 1 + \chi_0({\bf r})$;
we use units where $\epsilon_0=1$. The partition function of the fluctuating
dipoles
\begin{equation}
  Z= \int {\cal D}{\bf p}\; \exp{\left( -\beta U_p\right) },
\end{equation}
can be simplified by re-writing the Coulomb potential, $1/4 \pi r$ as an
integral over a potential $\phi$. We then find the effective energy
\begin{equation}
  U_{p, \phi} = \int \left( 
    {( \nabla \phi)^2 \over 2 } - i \phi \div {\bf p} +
    {{\bf p}^2({\bf r}) \over 2 \chi_0({\bf r})}
  \right)  \dV. \label{effective}
\end{equation}
One now performs the Gaussian integral over ${\bf p}$, to find the partition
function expressed as an integral over $\phi$. To do this it is useful to
integrate by parts replacing $ \phi \, \div \p$ with $- \p \cdot \grad \phi$;
boundary terms vanish in periodic systems or those in which fields decay at
infinity.

{In order to perform numerical calculations, we derive a discretized
  theory}.  We discretize by placing scalar quantities such as $\phi$ on
$V=L^3$ nodes of a cubic lattice. We choose a length scale such that the
lattice spacing is unity. The lattice spacing should be {much} larger than
the atomic scale, so that a formulation in terms of continuum dielectric
properties is possible; the lattice should however be sufficiently fine to
resolve features of physical interest, such as points or rough surfaces.
Vector fields, such as $\grad \phi$, are associated with the $3V$ links,
$\epsilon$ and $\chi_0$ are also associated with the links.  We find
\begin{equation}
  Z(\epsilon) = \prod_{l} \chi_{0,l}^{1/2} \int {\cal D} \phi\; 
  \exp \left (-\beta \sum_{l}
    {\epsilon_l \over 2}(\partial_l \phi)^2
  \right)
  \label{z2}
\end{equation}
where $l$ is a link; the discretization of the derivative $\partial_l$
evaluates the difference between variables on the corresponding nodes.  The
integral in eq.~(\ref{z2}) is over all modes excluding the uniform mode,
$\phi={\rm const}$ which has eigenvalue zero \cite{long}. 
We write the quadratic form
appearing in eq.~(\ref{z2}) as $\phi M \phi/2$ with a symmetric matrix $M$.
The interesting, long ranged, part of the free energy of interaction comes
from
\begin{equation} \label{detM}
  {\mathcal F}= {{k_BT}\over 2} \log{({\rm det }'(M))}
\label{free}
\end{equation} where the determinant again excludes the zero eigenvalue. 
We will see that the advantage of working with a matrix of the form
eq.~(\ref{z2}) rather than eq.~(\ref{marcus}) is that the quadratic form
eq.~(\ref{z2}) is sparse; a very
large majority of its elements are zero, enabling 
the use of efficient evaluation strategies.
While eq.~(\ref{marcus}) is {also} Gaussian, its quadratic form is dense.

Some care has to be exercised to separate various contributions to the free
energy eq.~(\ref{free}). In most situations of interest we want to find
interfacial energies, which are subdominant compared with the bulk free energy
of the full three dimensional system which we treat.  The problem of
separating bulk and surface contributions in general geometries is treated in
a separate publication \cite{long}.  In the first physical example presented
in this paper the volume of each phase changes as the tip is moved farther
from the indentation, and the surface free energy is obtained after
subtracting the bulk free energy.  In the second system we analyze, the volume
of each phase is constant so that variations in the interfacial free energies
can be found by simple subtraction, as the bulk contribution cancels out
completely.

\section{Matrix Factorization}
In the simplest geometries, such as multiple parallel plates, the determinant
is evaluated analytically from the eigenvalue equations for sets of plane
waves \cite{NinhamJCP}. Since we wish to work in arbitrary geometries we
evaluate the determinant eq.~(\ref{detM}) using general matrix methods, with
periodic boundary conditions to minimize edge effects.  Despite the fact that
$M$ is sparse, and can be stored in memory which is proportional to the volume
$V=L^3$, standard (dense) matrix methods find the eigenvalues in a time which
varies as $\tau_e \sim V^3$, using a storage $S\sim V^2$. Both these scalings
turn out to be prohibitive, and limit us to the study of systems no larger
than linear dimension $L=25$.  We thus turn to methods which are adapted to
large sparse matrices. For such matrices it is preferable to evaluate the
determinant without forming intermediate dense matrices, saving computer
memory and accelerating calculations.  We evaluate the determinant by firstly
modifying the matrix in order to render it positive definite, so that we are
not troubled by the zero eigenvalue. We do this by adding $V$ to a {\it
  single}\/ arbitrary diagonal element of the matrix; this can be shown to be
equivalent to neglecting the zero mode of the determinant \cite{long}.  With
the matrix, $M$ now positive definite we write it as a product of Cholesky
factors $M=R^T R$ where $R$ is an upper triangular matrix.  Because $R$ is
triangular the determinant is given by the product of the diagonal elements.
From the factor $R$ we find $\det M= (\det R)^2$.

This factorization has a number of remarkable properties that make it far more
powerful than diagonalization.  A good choice for the ordering that is used
for evaluating the Cholesky factors renders the method particularly
interesting: Nested dissection numbers the nodes in a non-consecutive manner
by recursively cutting a graph into equal pieces \cite{George}. For this
ordering of our matrix the Cholesky factor (in three dimensions) can be
generated with a storage of $S\sim V^{4/3}$, and with arithmetic effort
$\tau_e \sim V^2$ \cite{Duff}. We used the software package Taucs
\cite{Toledo} to perform the Cholesky factoring. We find that a system of
$V=64^3$ can be factored on a 3.2GHz 64-bit Xeon workstation in approximately
$300$ seconds, using 3GB of memory.

\section{Interactions between a tip and indentation} 
\label{approximation}

We now consider the interactions between a tip and a surface indentation in a
system in which the thermal contribution is dominant. Experimentally this can
occur either with tips in water which are made with materials which have
similar optical properties (so that the high frequency contributions to the
Lifshitz energy cancel out), or in systems that are sufficiently separated
such that the quantum interactions have died out requiring separations larger
than $\hbar c/2 k_B T\sim 5\mu m$ at room temperature \cite{Cappella,Daicic}.

\begin{figure}[ht]
  \includegraphics[scale=.9,angle=0] {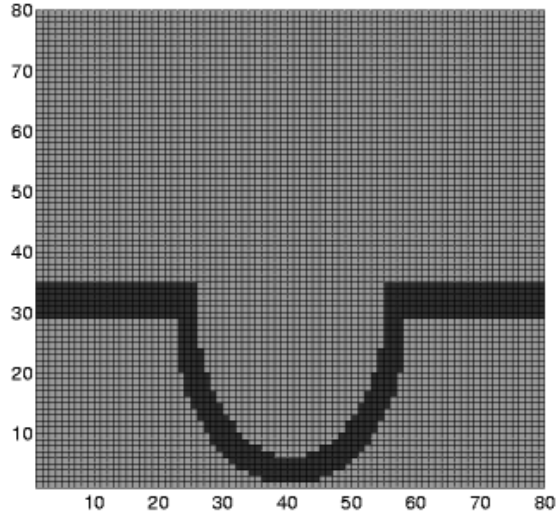}
  \caption{Section of the tip-indentation system taken on a plane going
    through the center of the tip. The system is discretized to a lattice with
    $L=80$. The tip and the indentation are taken to be half ellipsoids with
    vertical semi-axes $a=18$ and horizontal semi-axes $b=15$ and $b=19$
    respectively. In the figure $l=6$.}
  \label{spheretip}
\end{figure}

We evaluated the free energy of interaction for a system composed of a three
dimensional rounded tip close to an indentation in a surface
discretized to a lattice of dimension $L=80^3$, Figure~\ref{spheretip}.  We
evaluated the free energies with two vertical displacements of $l=6$ and
$l=20$. The physically interesting case for optically matched systems has a
contrast ratio of approximately $r_\epsilon=\epsilon_1/\epsilon_2=50$ between
the two media.  However in order to study the evolution of the interactions
with contrast we worked with a minimum ratio of $r_\epsilon=1.05$ to a maximum
of $r_\epsilon = 100$.

In the absence of a general method to compute dispersion forces in arbitrary
geometries, two approximations are commonly used.  The first is the proximity
force approximation according to which the interaction between surfaces of any
geometry is found by assuming that the surfaces are locally flat. One then
sums an effective, local free energy of the form
\begin{equation}
U(h,A) ={A \over 12 \pi h^2} \label{hamaker}
\end{equation}
over the surfaces where $h$ is the local distance between bodies and $A$ is
the Hamaker constant.
\begin{figure}[ht]
  \includegraphics[scale=1,angle=0] {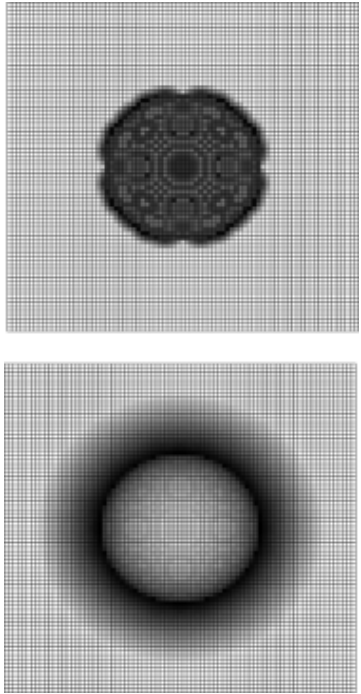}
  \caption{Minimal distance map for $l=6$ (top) and $l=20$ (bottom).  Black
    corresponds to the minimal distance, while white corresponds to the
    maximal distance. As the tip moves up the geometry of the closest approach
    changes from a central disk to a ring from the rim of the indentation.
    Mesh corresponds to the discretization lattice.}
  \label{minmap}
\end{figure}

\begin{figure}[ht]
  \includegraphics[scale=.4,angle=0] {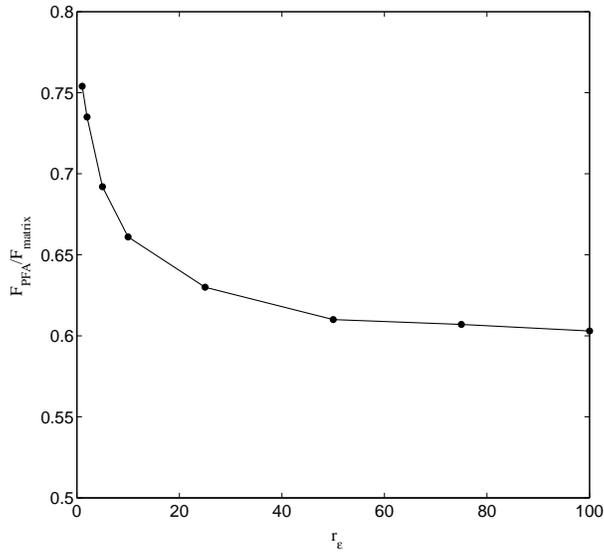}
  \caption{Ratio between the free energies computed via proximity force
    approximation and our numerical results as a function of the dielectric
    contrast between the two materials for  $l=6$.}
  \label{gap6}
\end{figure}

For the tip of Figure~\ref{spheretip} we compared results obtained via
proximity force approximation with our numerical results for ratios of the
dielectric constants varying from $r_\epsilon=1.05$ to $r_\epsilon = 100$.  In
order to construct the proximity force approximation, for each lattice site of
one surface we find the closest lattice site on the other surface and build a
minimal distance map.  We then use the generalization of eq.~(\ref{hamaker})
to periodic boundary conditions \cite{long} to evaluate the interaction at
each distance in the map and add all contributions to obtain the final free
energy.  The maps for the two chosen gaps are shown in Figure~\ref{minmap}.
We find that the free energies computed with the proximity force approximation
can deviate to up to $40\%$ from the full numerical evaluation. In particular
the proximity force approximation performs the worse for high dielectric
contrast ($r_\epsilon=50$) and when the two surfaces are close together.
Results for this system are shown in Figure~\ref{gap6}.  When the $l=20$
proximity force approximation performs better with deviations from the full
evaluation of $8\%$ for large contrast.

The second method, widely used in computer modeling, is an additivity
assumption according to which the overall interaction is found by summing
pairwise interactions of elementary components.  A typical example of such
approximation performs an integration of the long-ranged tail of Lennard-Jones
$1/r^6$ potentials between infinitesimal constituents of the interacting
macroscopic objects \cite{AFM-tips}.  More generally, if one assumes a
pairwise interaction between elements of a dielectric of the form $V(r) =
\alpha v(r)$ where $\alpha$ characterizes the strength of the potential, then
the long-ranged part of the interaction energy between two macroscopic bodies
can be written as
\begin{equation}
  U ({\bf R}, \alpha) = \alpha G({\bf R}) \label{pair2}
\end{equation}
 where the coordinates
${\bf R}$ are the relative positions of the bodies and the function $G$
encodes all the geometric information.

We notice that both eq.~(\ref{hamaker}) and eq.~(\ref{pair2}) display a
factorization property (they are expressed as a product of material and of
geometric terms) so that if we for instance consider the ratio
\begin{equation}
  {\mathcal R}_{12}=U ({\bf R}_1, \alpha) / U ({\bf R}_2, \alpha) \label{ratio}
\end{equation}
for two different geometries we find a result independent of the material
property.  Since this factorization property is true in two very different
limiting approximations it seems interesting to study the degree to which it
remains valid over a wide range of dielectric contrasts in a geometry which is
far from planar.

\begin{figure}[ht]
  \includegraphics[scale=.3,angle=0] {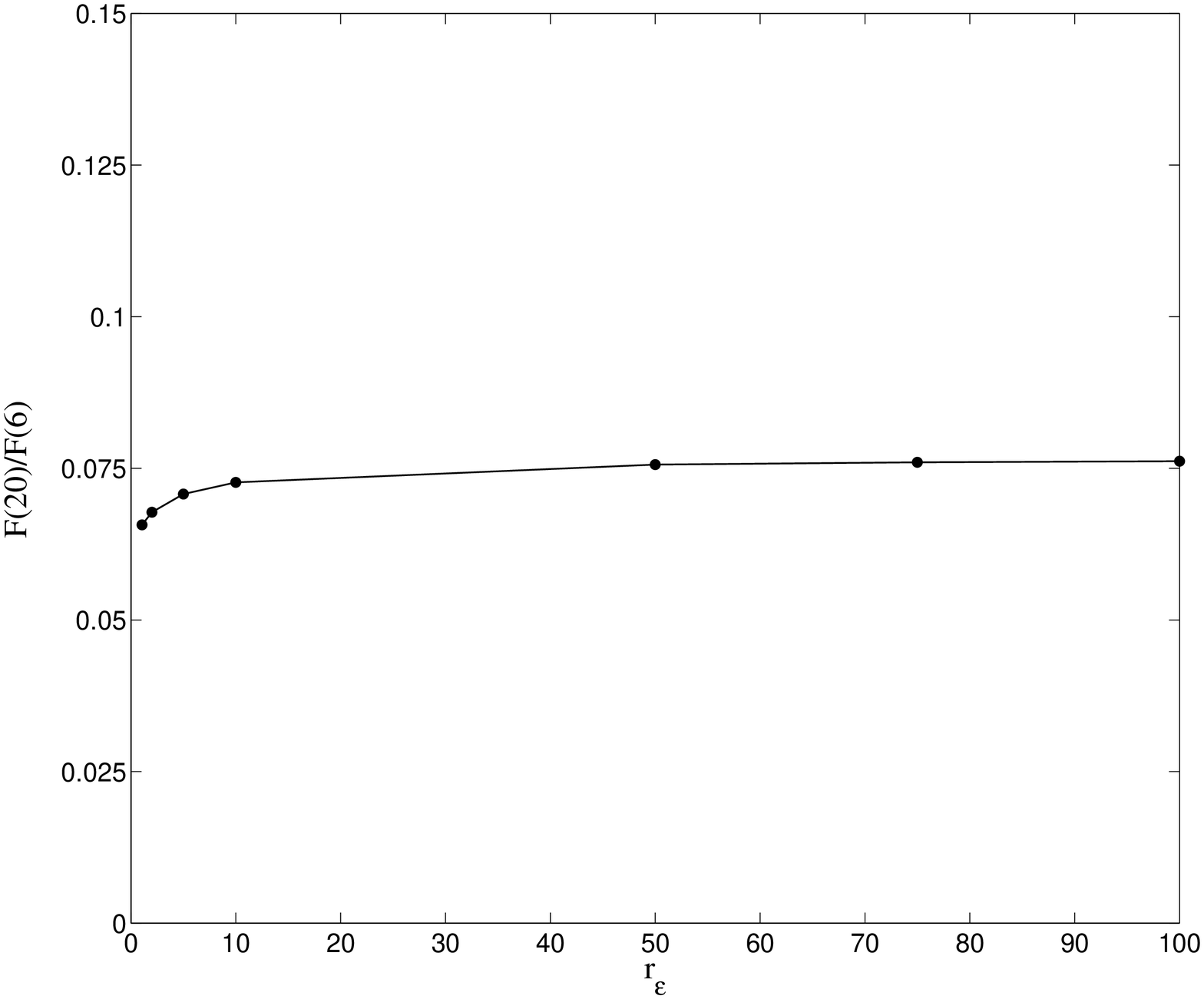}
  \caption{Test of the factorization property, eq.~(\ref{ratio}).  {The
      ratio of interaction energies for two separations of the tip-indentation
      system of Figure~\ref{spheretip} is plotted as a function of
      $r_\epsilon$}. The ratio varies by just 13\% over the whole range of
    dielectric contrasts, despite the great variation in geometry displayed in
    Figure~\ref{minmap}.  }
\label{factor}
\end{figure}

Surprisingly, Figure~\ref{factor}, we find that the ratio of free
  energies computed at different distances exhibits only a moderate variation
  with the dielectric contrast.  The robustness of this result has been tested
  by changing the system size and the accuracy of the tip/indentation
  discretization.  Our study suggests a way of inferring free energy values at
  high dielectric contrast once the results at low contrast are known.  It is
  sufficient to determine the interaction at low $\epsilon$ to obtain the
  whole set of measurements for the desired system.  Our results show that for
  classical, thermal interaction, the factorization method we propose, works
  better than proximity force approximation when large surface deformations
  are present and for high dielectric contrasts, both instances where
  proximity force approximation performs poorly.  This approach can be
  combined with analytic approximations that compute free energies for
  arbitrary geometries in a small dielectric contrast expansion
  \cite{Golestanian}.

\section{Non-retarded quantum interactions} 
\label{non-retarded}

The quantum interaction between two materials is classified as non-retarded
when the interaction is instantaneous, or retarded when one must take into
account the finite propagation speed of electromagnetic radiation.  In the
first case the decay of interactions between two atoms is $1/r^6$; in the
second the interaction falls as $1/r^7$. The characteristic crossover distance
between these two forms is determined by the wavelength of typical spectral
features that dominate dielectric response. This is often a feature in the
ultraviolet, leading to a cross-over length of tens of nanometers for most
materials.

We now generalize our treatment of the interaction to the short
distance, non-retarded regime. The systems for which this effect is dominant
are limited to the nano-scale, however we will show that this case is
particularly simple to treat with the methods developed above. We leave the
generalization to retarded interactions, which require the correct treatment
of the propagating electromagnetic field, for a future publication.

\subsection{Quantization}
Since in the energy eq.~(\ref{marcus}) the field $\p$ represents a microscopic
polarization vector we can study its dynamics by adding the kinetic energy.
\begin{equation}
  T=\rho({\bf r }) \dot {\bf p}^2/2
\end{equation}  
where $\rho({\bf r})$ is a mass density.

The thermodynamics of a quantum system are  particularly simple to treat
with the method of path integral quantization \cite{Ceperley}. The potential
and kinetic energies are combined in an effective statistical weight for an
ensemble of $N$ identical replicas of the original system; these replicas are
coupled in the time/temperature direction by harmonic springs. The exact
quantum partition function is then generated in the limit of large $N$.

The effective action at each time slice, $m$ is
\begin{equation}
  U_m =\int \dV {\rho({\bf r})( {\bf p}_m - {\bf p}_{m+1})^2 
    \over
    2 \hbar^2 \tau} + \tau U_{p} \label{springs}
\end{equation}
with $\tau= \beta/N$. The harmonic coupling between the replicas comes from
the presence of the terms in $m$ and $m+1$ in eq.~(\ref{springs}) arising
from the kinetic energy.  We now perform a Fourier transform in the
time/temperature direction and find that the energy can be written in a form
which is identical to eq.~(\ref{marcus}) if we define a frequency dependent
susceptibility.
\[
{1\over \chi(\omega)} =  {1\over \chi_0} + 
{\rho \over \hbar^2 \tau^2} (2 -2 \cos\omega) 
\]
with Fourier frequency $\omega=2 \pi n/N$.

We again perform the transformation from eq.~(\ref{marcus}) to
eq.~(\ref{effective}) by introducing an integral over the potential. After
integrating over $\p$ the partition function of the quantum system is given by
a product
\begin{equation}
  Z= \prod_{n=0}^{N-1} Z( \epsilon(n) ) \label{zn}
\end{equation}
where in the $Z(\epsilon)$ of eq.~(\ref{z2}), the  dielectric function
is replaced by 
\begin{equation}
  \epsilon(n) = 1 + 
  {\chi_0 \over 1 + 
    N^2 (2-2 \cos(2 \pi n/N ))/\hbar^2 \omega_0^2 \beta^2 
  }
  \label{omega}
\end{equation}
with $\omega_0= 1/\sqrt{\chi_0({\bf r}) \rho({\bf r})}$.  If we introduce the
Matsubara frequencies $\omega_n=2n\pi/\beta$ we find that for $N$ large
eq.~(\ref{omega}) simplifies to the single pole approximation
\begin{equation}
  \epsilon(n) = 1 + {\chi_0\over  1 + \omega_n^2/\hbar^2\omega_0^2} \label{pole}
\end{equation}
often used \cite{Israelashvili} to fit the dielectric properties of material in
calculations of the Hamaker constant.  
Passage to the limit eq.~(\ref{pole}) requires that
\begin{equation}
  N^2 \gg (\hbar \omega_0   \beta)^2 {\epsilon(0)} \label{crit}
\end{equation}
in order that the high frequency limit $\epsilon(\omega) \rightarrow 1$ is
correctly reproduced.  

This path integral formulation gives the full, combined thermal and quantum
contribution to the interaction potential. It does, however, become less
efficient at low temperatures; at exactly zero temperature the free energy 
should be evaluated by direct numerical quadrature over frequency.

\subsection{Implementation}

Each contribution to eq.~(\ref{zn}) at frequency $n$ requires the evaluation
of a single matrix determinant identical in form to that evaluated in the
classical interaction.  We measured the absolute discretizion error by
studying the free energy of interaction between two parallel slabs for
different values of $N$.  We followed the procedure of \cite{long} using
$\epsilon(0)=5$, $\hbar \omega_0 \beta =40$. If the largest contributions to
the free energy comes from the ultra-violet range of the spectrum our choice
of the parameters corresponds to a system near room temperature. Quantum
effects then dominate the interaction, but thermal effects are correctly
included in the evaluation.

\begin{figure}[ht]
  \includegraphics[scale=.4,angle=0] {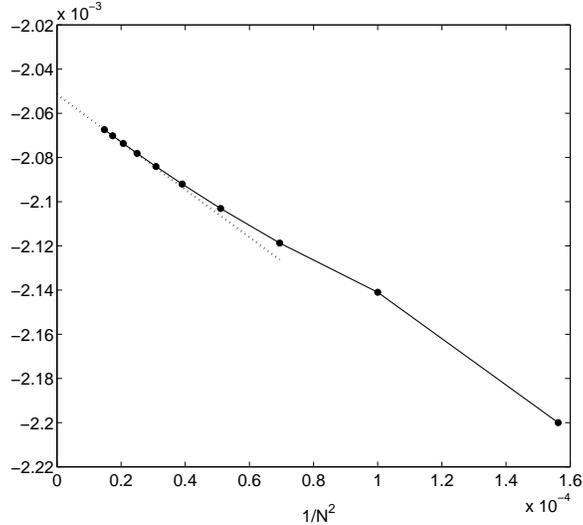}
  \caption{Convergence of free energy differences with copy number $N$. We
    evaluated the free energy difference of two parallel plates separated by
    $l=2$ and $l=10$ for $N$ varying from $80$ to $260$.  The free energy is
    plotted as a function of $N^{-2}$. From values of $N$ above $180$ we
    extrapolate $F_{\infty}= 2.051 \times 10^{-3}$.  L=48.}
  \label{richardson}
\end{figure}

From eq.~(\ref{crit}) we require that $N\gg 90$.  We performed evaluations of
the free energy for $80 \le N \le 260$, Figure~\ref{richardson}.  Beyond
$N=180$ the free energy can be fitted by the Richardson form $\mathcal{F} =
\mathcal{F}_{\infty} + b/N^2$ \cite{Brualla}, where $\mathcal{F}_{\infty}$ and
$b$ are fitting parameters.  Using values of $N$ greater than $180$ we
estimate $\mathcal{F}_\infty$.  We then evaluate the error generated for a
specific value of $N$.  We find an agreement that varies from $93\%$ for
$N=80$ to $99.0\%$ for $N=260$.  We adopted $N=240$ for the evaluation of the
free energy landscape described below.  For this value $\mathcal{F}_\infty
/\mathcal{F}_{240} = 0.987$.  Higher accuracy and faster convergence (in
$1/N^4$) are possible if we work with two values of $N$ and extrapolate to
$\mathcal{F}_{\infty}$.

We then evaluated the free energy of a system composed of a sharp point over a
surface with regular wells as a function of the horizontal position.  Given
that typical small force microscopy tip sizes are of the order of $10nm -
50nm$ forces measured experimentally by this technique will generally fall into
the crossover between the retarded and non-retarded regimes.  However, there
do exist extreme cases of tips of radii of $2nm$ (super sharp silicon tips),
where this case could be relevant, even if we are beginning to be close to the
atomic scale where a continuum dielectric description is more difficult to
justify.

\begin{figure}[ht]
  \includegraphics[scale=1,angle=0] {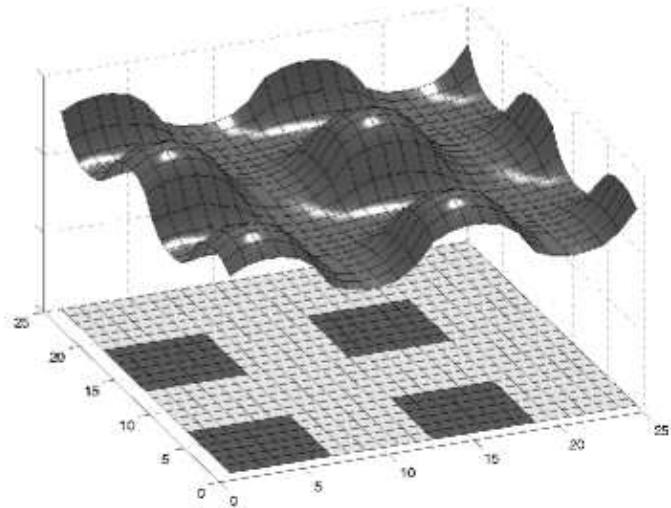}
  \caption{Quantum free energy landscape for a system composed of a sharp,
    conical, tip near a surface with regular wells, side and depth $l=6$,
    separated by $6$ lattice spaces in both $x$ and $y$ direction. Box size
    $L=48$.  A portion of the system of dimensions $12\times 12$ is scanned
    and the result reproduced periodically.  Top surface: the free energy
    landscape at one lattice space above the top of the wells.  Lower surface
    indicates the positions of the wells.}
  \label{qland}
\end{figure}

We scanned a region of $12 \times 12$ lattice points, covering the area of one
well and its surroundings.  Results of the free energy landscape are shown in
Figure~\ref{qland}.  The free energy of each point $\mathcal{F}(x,y)$ is
evaluated for $N=240$, for which, due to the symmetry
$\epsilon(n)=\epsilon(N-n)$, only $N/2+1=121$ determinants are needed.
Firstly, the matrices ($12^2 \times 121 \sim 17,500$) were built using Matlab,
and stored on disk.  This took approximately two days on a single workstation.
Then, the free energy evaluation was performed on a cluster of nine
processors, and took approximately 2 days to complete, with the evaluation of
the interaction at each position taking about 3 hours.  The energy landscape
we obtained reflects the underlying well profile, but with smoothed features.

\section{Discussion}
Quite {fine}
%large 
and useful discretization of fluctuating dielectrics can be
studied using methods based on Cholesky factorization of a sparse determinant.
One can study the full multibody dispersion interaction at nanometric length
scales, a scale suitable for the study of macromolecules and nano-particles,
in non-trivial three dimensional geometries.  Conventional methods for
interpreting such systems use extensive modeling with force fields that
contain many simplifying assumptions which can now be checked (in the near
field regime) against explicit numerical results.

We now compare our approach with recent numerical work \cite{MIT}. The authors
work with the full discretized Maxwell equations which enables them to
consider the full retarded interaction between bodies; they  calculate
the stress tensor rather than the free energy that we chose to evaluate. The
method requires the calculation of Green functions which are then integrated
over a closed surface surrounding the body of interest. The authors propose
several different methods of solving for the Green functions, including fast
multipole and multigrid methods, which they argue can solve the problem in
with memory $S\sim V$ and time $\tau_e \sim V^{2-1/d}$. They work with modest
system sizes of $20\times40$ in simplified $2+1$ dimensional geometries,
suitable for studing grooved surfaces.

In practice they used a conjugate method which requires a number of iterations
which increase with the systems size as $V^{1/d}$ \cite{golub}, so that their
actual implementation scales like our own as $\tau_e \sim V^2$.  It could have
been useful to compare real computing times, and not only theoretically
derived asymptotic behaviors, but unfortunately the authors do not report
these data.  Any consideration of the choice of algorithm must take into
account the prefactors in these laws; fast multipole methods have been
abandoned in applications such as molecular dynamics due to the enormous
prefactors in the (apparently favorable) asymptotic scaling.

For system sizes comparable to those used in this paper one finds that the
number of floating point operations needed to perform the Cholesky
factorization is comparable to $N_{flop}= 6.5 V^2$ \cite{Duff2}.  Multigrid
methods while being asymptotically fast can require hundreds of
iterations in order to converge \cite{roberts}. It thus seems possible that they perform less
well than the methods of the present paper for moderate system sizes. It would
be most interesting to study the crossover point in the efficiency of the
various methods.

When we restrict ourselves to systems in $2 +1$ dimensions the Cholesky
factorization we use also improves in performance.  Storage requirement in
this case is only $S\sim N\log (N)$, and computation time is $\tau_e\sim
N^{4/3}$.  The interaction of grooved surfaces is studied by Fourier
transforming in the uniform direction before performing the factorization \cite{long}. 
In such a system evaluation of the interaction of a system discretized to a
lattice of dimensions $500^3$ can be performed in $30$ minutes {on a $3.2$GHz $64$-bit 
Xenon workstation}.

The main physical problems that are still not possible to study with this
method have length scales in the range $20nm - 10\mu m$ where the retarded
interaction dominates.  This requires a different discretization strategy
based on the full Maxwell equations and will be considered in an future paper
\cite{upcoming}.

We wish to thank F. Nitti for many useful discussions.  Work financed in part
by the Volkswagenstiftung.

\bibliography{thermalC_general}
\end{document}